\documentclass[a4paper,11pt]{article}
\pdfoutput=1 

\usepackage{jheppub} 

\usepackage[T1]{fontenc} 
\usepackage[utf8]{inputenc}
\newcommand\f[2]{\frac{#1}{#2}} 
\newcommand\as{\alpha_{\mathrm{S}}} 
\def\beq{\begin{equation}} 
\def\eeq{\end{equation}} 
\def\to{\rightarrow} 
\def\beeq{\begin{eqnarray}}
\def\eeeq{\end{eqnarray}}

\title{\boldmath Higgs boson pair production at NNLO in QCD including dimension 6 operators }

\author[a]{Daniel de Florian,}
\author[a]{Ignacio Fabre,}
\author[b]{and Javier Mazzitelli}

\affiliation[a]{International Center for Advanced Studies (ICAS), ECyT-UNSAM,\\Campus Miguelete, 25 de Mayo y Francia, (1650) Buenos Aires, Argentina}
\affiliation[b]{Physik-Institut, Universit\"at Z\"urich,\\Winterthurerstrasse 190, CH-8057 Z\"urich, Switzerland}

\emailAdd{deflo@unsam.edu.ar}
\emailAdd{ifabre@unsam.edu.ar}
\emailAdd{jmazzi@physik.uzh.ch}

\abstract{In this paper we present the computation of the Higgs boson pair production cross section, both inclusive as well as differential on the invariant mass distribution, at next-to-next-to-leading order (NNLO) in QCD including effects of new physics beyond the standard model. We parametrize the effects of new physics with the relevant dimension 6 operators in a standard model effective field theory (EFT) approach, and examine their phenomenology.
The dependence of the NNLO $K$-factor on the EFT couplings is analysed, finding that, while rather flat for a number of EFT coefficients, it can considerable differ from the  standard model value in some particular regions of the parameter space.
We present explicit examples of (almost) degeneracy in the NNLO cross-section with respect to the anomalous couplings, showing that the invariant mass distribution has a much larger sensitivity to phenomena beyond the standard model and can be used as a tool to  discriminate their effect.}

\begin{document} 
\maketitle
\flushbottom

\section{Introduction}

Since the discovery of a scalar boson in 2012~\cite{Aad:2012tfa,Chatrchyan:2012xdj,Khachatryan:2016vau}, the high energy physics community is focused in determining whether it corresponds to the long sought standard model (SM) Higgs boson~\cite{Higgs:1964ia,Higgs:1964pj,Englert:1964et}, or if it might provide the first hint for new physics beyond the SM (BSM). 

Great  improvement was achieved during the last few years, both from the theoretical side --by providing more precise results for a number of observables-- and from the experimental measurements, in order to extract the couplings between the Higgs boson and the 
third generation of quarks and leptons.
Besides these important results, it is also crucial to study the Higgs self-couplings, which provide a way to explore the potential that thrives electroweak symmetry breaking. 
In the SM, the Higgs self-couplings are uniquely fixed via enforcing the preservation of the corresponding gauge symmetries and renormalizability~\cite{tHooft:1971qjg,tHooft:1972tcz}, and any deviation would be a sign of BSM physics.

The production of Higgs boson pairs provides a direct way to test the Higgs trilinear coupling (see also Refs.~\cite{Degrassi:2016wml,Bizon:2016wgr,Degrassi:2017ucl} for an alternative approach), and as it happens for the single Higgs boson production cross section, the dominant production channel proceeds at hadron colliders via gluon fusion, mediated by heavy-quark loops~\cite{Glover:1987nx,Eboli:1987dy,Plehn:1996wb}. 
This means that BSM physics can be realized in two distinct ways, either through a resonance due to a reasonably light field that acts as a mediator, or through heavier fields that participate in the loop thus modifying the effective couplings between the SM particles. 
In the former, a direct detection can be achieved by looking at the invariant mass spectrum of the final state, but in the latter a precision measurement has to be performed in order to search for deviations from the SM. 
A model-independent approach to parametrize BSM effects consists in considering the low energy effective field theory (EFT) that remains after integrating out the heavy fields of new physics (NP), introducing higher dimensional operators suppressed by the mass of these heavy particles ($\Lambda$). In principle one could consider to add all possible higher order operators to the SM Lagrangian that are compatible with its symmetries, up to some power of $\Lambda$~\cite{Burges:1983zg,Leung:1984ni,Buchmuller:1985jz,Grzadkowski:2010es}. 
Expanding up to dimension 6 ($\Lambda^{-2}$), 2499 of such operators are found~\cite{Alonso:2013hga}. However, if we consider only those that vanish in the absence of the Higgs boson, and therefore, those that only contribute to observables when at least one Higgs boson participates in the process,
the number reduces to $5$, which can be written in several different basis.

Given that the leading order (LO) contribution to Higgs pair production occurs at loop level, higher orders in QCD perturbation theory are extremely difficult to calculate. Recently, a complete next-to-leading order (NLO) computation became available~\cite{Borowka:2016ehy} that evaluates numerically the integrals of the required multi-scale two-loop amplitudes. In the heavy-top limit (HTL), where the top quark is considered heavy and the rest massless, NLO corrections have been presented in Ref.~\cite{Dawson:1998py} and a rescaling with the exact Born cross section was performed. The NLO corrections represent an increase of about $100\%$, and the HTL result is only about $14\%$ bigger than the exact result from Ref.~\cite{Borowka:2016ehy}. While the computation of the three-loop virtual corrections is presently out of reach, working within the HTL it is possible to compute corrections beyond NLO, like the ones derived in Refs.~\cite{deFlorian:2013uza,Grigo:2014jma}, which allowed for the next-to-next-to-leading order (NNLO) result presented in Ref.~\cite{deFlorian:2013jea}. This calculation allows to compute not only the inclusive cross section (resulting in an increase of about $20\%$ at this order), but also the differential invariant mass distribution of the produced pair of Higgs bosons. Within the same heavy-top limit, a fully exclusive calculation at NNLO was recently presented in Ref.~\cite{deFlorian:2016uhr}.
With respect to computations that include dimension 6 operators in the SM EFT approach~\cite{Goertz:2014qta}, bounds for the Wilson coefficients of the new operators have been obtained from the data collected by the LHC and earlier experiments by different groups~\cite{Pomarol:2013zra,Falkowski:2014tna,Ellis:2014jta,Dumont:2013wma,Falkowski:2015fla,Butter:2016cvz}, and a NLO calculation for Higgs pair production in the HTL was performed in Ref.~\cite{Grober:2015cwa}.

In this work we present the computation of the Higgs bosons pair production cross section at NNLO in QCD, including the relevant dimension 6 operators of the SM EFT. The paper is organized as follows: In section~\ref{sec:Details} we provide the details of the calculation, including  two popular basis for the EFT. In section~\ref{sec:pheno} we present the phenomenological results, discussing the dependence of the $K$-factor  on the higher dimensional couplings and the degeneracy of the inclusive cross section on them.
Finally, in section~\ref{sec:conc} we present our conclusions.

\section{Details of the calculation}
\label{sec:Details}

\subsection{EFT basis}
\label{sec:eft}
Heavy states of NP can be integrated out to obtain a low energy effective Lagrangian, with new contact interactions that would be otherwise mediated by the NP states. The coupling constant of the effective  interaction is therefore suppressed by the mass scale $\Lambda$ of the new heavy state, making the operator of dimension higher than four, and therefore non-renormalizable.
In this context, in order to parametrize all possible BSM theories that are free of new light states, one should include in the SM Lagrangian all higher dimensional terms that are consistent with the SM symmetries. There are 1350 CP-even and 1149 CP-odd of such dimension 6 (${\cal O}(\Lambda^{-2})$) operators~\cite{Alonso:2013hga}. Nevertheless, we will focus only on the operators that vanish in the absence of the scalar boson, as the others can be better constrained by other observables. 
This includes for instance the chromomagnetic operator, that introduces a coupling between gluons, the top quark and the Higgs doublet, which in recent studies~\cite{Maltoni:2016yxb} has been shown to mix with other operators. 
Nevertheless, the effects arising from this operator also appear in the absence of the Higgs boson (from the term proportional to the Higgs vacuum expectation value) and therefore introduce corrections to {\it pure} QCD processes, in particular to the top pair production cross section that can  set a better constraint on the corresponding anomalous coupling~\cite{Franzosi:2015osa}.
Formally, this operator introduces corrections of order $\mathcal{O}(y_t^2)$ to the results of this work, where $y_t$ is the top Yukawa coupling, and therefore is not considered herein.

 In this work we will neglect the mass of all fermions (thus their couplings to the Higgs) except for the top quark, since their contributions to Higgs pair production through gluon fusion accounts for less than $1\%$ of the LO cross section in the SM \cite{Baglio:2012np}. Due to this suppression, these contributions were not considered in the present work, although in principle they could be enhanced in some BSM scenarios.

 If one considers the Higgs boson $h$ as a singlet of the custodial symmetry, and not necessarily part of an $SU(2)_L$ doublet, the relevant dimension 6 operators can be written as~\cite{Contino:2010mh}
\beeq\label{eq:Lagrangian}
{\cal L}_\text{non-lin} \supset &-& M_t \, \bar{t} t \left(c_t  \frac{h}{v} + c_{tt} \frac{h^2}{2 v^2}\right) - c_3 \frac{1}{6}\left(\frac{3M_h^3}{v}\right) h^3 \nonumber \\
&+& \frac{\alpha_s}{\pi} G^{a\,\mu \nu} G^a_{\mu \nu} \left(c_g \frac{h}{v} + c_{gg} \frac{h^2}{2 v^2}\right)\,,
\eeeq
where $\alpha_s=g_s^2/(4 \pi)$, $g_s$ is the strong coupling constant, $t$ represents the top quark with mass $M_t$, $v\approx 246$~GeV is the SM Higgs field vacuum expectation value, $M_h$ is the mass of the Higgs boson $h$, $G^a_{\mu \nu}$ is the gluon field strength tensor, and $c_{i=t,tt,3,g,gg}$ are the Wilson coefficients, after a canonical normalization of the Lagrangian. The operators parametrized by $c_t$ and $c_3$ modify the ones already present in the SM, namely the coupling between the Higgs boson and the top, and the Higgs boson self-coupling, respectively. The rest of the operators are new, $c_g$ and $c_{gg}$ parametrizing the contact interaction between gluons and one and two Higgs bosons, respectively, and $c_{tt}$ the one between the top quark and a pair of Higgs bosons. 
In this Lagrangian, the $SU(2)_L\times U(1)_Y$ symmetry is non-linearly realized, and the SM corresponds to the point in parameter space given by $c_3 = c_t = 1$ and $c_{tt}=c_g=c_{gg} = 0$.

Using a different approach, one could assume that the Higgs boson is part of an $SU(2)_L$ doublet $H$.
This particular extension of the SM is included in the so-called SILH basis~\cite{Giudice:2007fh}, and is given by the operators
\beeq\label{eq:SILH}
{\cal L}^\text{SILH}_6 &\supset& \frac{\bar c_H}{2 v^2} \partial_\mu(H^\dagger H)\partial^\mu(H^\dagger H)
+\frac{\bar c_u}{ v^2} y_t(H^\dagger H \bar q_L H^c t_R + h.c.)\nonumber\\
& &-\frac{\bar c_6}{6 v^2} \frac{3 M_h^2}{v^2} (H^\dagger H)^3
 + \bar c_g \frac{g_s^2}{M^2_W} H^\dagger H G^{a\,\mu \nu} G^a_{\mu \nu},
\eeeq
where $M_W$ is the mass of the $W$ boson, and $\bar c_{i=H,u,6,g}$ are 4 free parameters (the SM corresponding to $\bar c_i=0$ for all $i$).
Expanding $H$ around $v$ in the physical gauge, one finds that it corresponds to ${\cal L}_\text{non-lin}$ with
\beeq
c_t&=&1-\frac{\bar c_H}{2} - \bar c_u,\qquad \;
c_{tt}=-\frac{1}{2}(\bar c_H + 3 \bar c_u),\nonumber \\
c_3&=&1-\frac{3}{2}\bar c_H+\bar c_6,\qquad
c_g=c_{gg}=\bar c_g \left(\frac{4\pi^2 v^2}{M^2_W}\right).
\eeeq
Bounds for the SILH coefficients can be found in Ref.~\cite{Contino:2013kra}. In this work we will use the more general extension of the SM, ${\cal L}_\text{non-lin}$.

\subsection{NNLO results}
\label{sec:calc}

Performing higher order QCD calculations for Higgs boson pair production has been probed to be quite complicated. Even within the SM, the NLO computation could be obtained only very recently~\cite{Borowka:2016ehy}. Therefore, an approach that is widely used consists in integrating out the top quark of the SM Lagrangian and work with the remaining effective theory that is valid for energy scales smaller than $\sim 2 M_t$, in what is called the heavy-top limit (HTL).

In recent higher order calculations~\cite{deFlorian:2013jea}, the QCD corrections were computed in the HTL, and then multiplied by the LO exact result, providing a rescaling (known as \emph{Born-improved HTL} or simply \emph{B-i. HTL}) that improves the accuracy of this approach~\cite{Dawson:1998py}. The exact NLO order calculation has shown that --despite the fact that the bulk of the cross section comes from di-Higgs invariant masses larger than the threshold $2M_t$-- this procedure is a rather good approximation at NLO, being the exact result for the inclusive cross section at NLO only a $14\%$ smaller than the B-i. HTL one~\cite{Borowka:2016ypz}, to be compared to the radiative correction of about $100\%$. 
Of course, the situation is in general different for kinematic distributions, finding --for some of them-- large discrepancies between the full NLO and the aforementioned approximation (see Ref.~\cite{Borowka:2016ehy} for a detailed comparison).
However, the differences in the shape of the Higgs-pair invariant mass distribution, which is the only one considered in this work, are moderate, being always below $25\%$ in the whole mass range under analysis for the SM.
It is worth to mention, though, that the size of the NNLO corrections in the HTL is of the same magnitude of the top-mass effects at NLO. Therefore, an EFT analysis combining both the full NLO and the HTL NNLO corrections, which is beyond the scope of this work, is highly desirable.

In this work we follow a similar scheme as in Ref.~\cite{deFlorian:2013jea}: we compute the QCD corrections within the effective theory where the top quark has been integrated out, and then we rescale the result in such a way that the exact LO cross section is recovered.

After integrating the top quark field in Eq.~\eqref{eq:Lagrangian}, the effective Lagrangian reads
\beeq\label{eq:Leff}
{\cal L}_\text{non-lin}^\text{HTL} \supset & & \frac{\alpha_s}{\pi} G^{a\,\mu \nu} G^a_{\mu \nu}\left\lbrace  \frac{h}{v} \left[\frac{c_t}{12} C_H+ c_g \right] +\frac{h^2}{2 v^2} \left[ - \frac{c_t^2}{12} C_{HH} + \frac{c_{tt}}{12} C_H + c_{gg}\right] \right\rbrace \nonumber \\
&-& c_3 \frac{1}{6}\left(\frac{3M_h^3}{v}\right) h^3
,
\eeeq
where $C_H$ and $C_{HH}$ are the coefficients that arise from matching the heavy-top effective theory to the full theory at NNLO. These are given by~\cite{Kramer:1996iq,Chetyrkin:1997un,Grigo:2014jma,Spira:2016zna}
\beeq
C_H &=& 1 + \frac{11}{4} \frac{\alpha_s}{\pi} + \left(\frac{\alpha_s}{\pi}\right)^2 \left[ \frac{2777}{288} + \frac{19}{16} \log \frac{\mu^2_R}{M^2_t} + N_f \left(-\frac{67}{96} + \frac{1}{3} \log \frac{\mu^2_R}{M^2_t} \right) \right] \nonumber \\ & &+\, {\cal O}(\alpha_s^3)\,,\\
C_{HH} &=& C_H + \left(\frac{\alpha_s}{\pi}\right)^2 \Delta C^{(2)}_{HH}+ {\cal O}(\alpha_s^3)\,,\\
\Delta C^{(2)}_{HH} &=& \frac{35}{24} + \frac{2}{3} N_f \,.
\eeeq

We can see in Eq.~\eqref{eq:Leff} that the coefficients $c_i$ only modify the effective couplings between the Higgs and gluons present in the SM. This simplification provides a straightforward way of generalizing the SM result to the EFT that includes dimension 6 operators. 

We follow the approach described in Ref.~\cite{deFlorian:2013jea}, where we distinguish between (a) contributions to the squared matrix element that have only two effective vertices between Higgs boson(s) and gluons ($\sigma^a$), and (b) those with more than two effective vertices ($\sigma^b$). Thus the partonic cross section, differential in the invariant mass of the di-Higgs system $Q$, can be written as
\beq \label{eq:sigma_exp}
Q^2 \frac{{\rm d}\hat{\sigma}}{{\rm d}Q^2} = \hat{\sigma}^a + \hat{\sigma}^b,
\eeq
where $\hat\sigma^a$ receives the same corrections than those for single Higgs production, due to the similarity of the amplitudes involved. We found that for each partonic subprocess $ij\to HH + X$, and for factorization and renormalization scales $\mu_R=\mu_F=Q$, the result is
\beeq
\hat{\sigma}^a_{ij} &= & \hat{\sigma}_{\rm LO} \left[ \eta_{ij}^{(0)} + \left(\frac{\alpha_S}{2 \pi}\right) 2\eta_{ij}^{(1)}+ \left(\frac{\alpha_S}{2 \pi}\right)^2 4 \eta_{ij}^{(2)}\right] \nonumber \\
&  -& \int_{t_-}^{t_+}\!\!dt \f{G_F^2\, \as^2}{512(2\pi)^3} \; \left\lbrace  \left(\frac{\alpha_S}{2 \pi}\right) \delta_{ig}\,\delta_{jg} \delta(1-x) 4 \,C_H^{(1)} {\rm Re}(A^* C_{LO})\right. \nonumber\\
 &+& \left. \left(\frac{\alpha_S}{2 \pi}\right)^2\delta_{ig}\,\delta_{jg} \delta(1-x)\,4 
\left[ 2\,{\rm Re}(C_{LO}^* \, F_\square)  \,c_t^2 \, \Delta C^{(2)}_{HH} 
 -  (C_H^{(1)})^2 |A|^2 + C_H^{(2)}\,2\,{\rm Re}(A^* C_{LO}) \right]\right.\nonumber\\
& +& \left.  \left(\frac{\alpha_S}{2 \pi}\right)^2 8 {\rm Re}(A^* C_{LO})\, C_H^{(1)} \eta_{ij}^{(1)}\right\rbrace,\label{eq:sigmaA}
\eeeq
where the coefficients $\eta_{ij}$ are those expressed in Ref.~\cite{Anastasiou:2002yz} (which also agree with the results presented in Refs.~\cite{Harlander:2002wh,Ravindran:2003um}),  $A$ and $C_{LO}$ are defined as
\beeq
A &=& \f{2}{3}\left[c_3 C_\triangle\, 12 \,c_g + 12\, c_{gg}\right]\,,\\
C_{LO} &=& c_3 \,C_\triangle \left(c_t\, F_\triangle + \f{2}{3} 12 \,c_g\right) + c_t^2\, F_\square + c_{tt}\, F_\triangle + \f{2}{3} 12\, c_{gg} \label{eq:CLO} \,,
\eeeq
$F_\triangle$, $F_\square$ and $G_\square$ are the usual triangle and box form factors and can be found in Ref.~\cite{Plehn:1996wb}, $C_\triangle$ includes the Higgs propagator
\beq\label{eq:Ctriangulo}
C_\triangle = \frac{3 M_H^2}{Q^2-M_H^2 + i M_H \Gamma_H}\,
\eeq
where $\Gamma_H$ is the Higgs total width, and $\hat{\sigma}_{\rm LO}$ is the LO partonic cross section for $gg\to HH$
\beq
\hat\sigma_{\text{LO}}= \int_{t_-}^{t_+}\!\!dt
\f{G_F^2\, \as^2}{512(2\pi)^3}
\left\{
\left|C_{LO}\right|^2
+ \left| c_t^2 G_\square \right|^2
\right\}\,.
\eeq
The integration variable is 
\beq\label{eq:t}
t=-\f{1}{2}\left(Q^2-2M_H^2- Q\sqrt{Q^2-4M_H^2}\cos\theta_1\right)\,,
\eeq
where $\theta_1$ is the scattering angle in the Higgs center-of-mass system. The limits of integration correspond to $t_{\pm} = t\,(\cos\theta_1 = \pm 1)$.

The effective vertex between gluons and Higgs is proportional to $\alpha_s$, which implies that $\hat \sigma^b$ is NLO at tree-level, and at NNLO there are  one-loop virtual and single real emission corrections. In the SM result from Ref.~\cite{deFlorian:2013uza}, $\sigma^b$ is split into different pieces, making it possible to keep track of the different amplitudes contributing to each of them, and thus to adapt this result to our EFT by inserting the appropriate factors. The renormalized result (for $\mu_F=\mu_R=Q$) can be written for the different channels as
\beeq
\hat \sigma^b_{gg} &=& \hat \sigma^{(sv)}_{gg} + \, \left(\hat \sigma^{(c+)}_{gg} +\hat \sigma^{(c-)}_{gg} +\hat \sigma^{(f)}_{gg}  \right),\label{eq:sBgg}\\
\hat\sigma^{(sv)}_{gg}&=& 
\int_{t_-}^{t_+}\!\!dt \f{G_F^2\, \as^2}{512(2\pi)^3}\, \delta(1-x)
\bigg\{
\left(\f{\as}{2\pi}\right)\f{4}{3}\,\text{Re}\left(C_{LO}^*\,V_\text{eff}^2 \right)
\nonumber
\\
&+&\left(\f{\as}{2\pi}\right)^2\bigg[
\text{Re}\left(C_{LO}^*\,V_\text{eff}^2 \right)\, \left(
\f{8\pi^2}{3}+{\cal R}^{(2)}-8\,\Delta C^{(2)}_{HH}
\right)
+\text{Im}\left(C_{LO}^*\,V_\text{eff}^2 \right)\,{\cal I}^{(2)}
\nonumber\\
&+&\left|V_\text{eff}\right|^4\, {\cal V}^{(2)}
- \f{22}{3}\,\left( 2\, {\rm Re}\left(C_{LO}^* V_\text{eff}\right)\, \f{2}{3}\, 12\,c_g + \,{\rm Re}\left(A^* V_\text{eff}^2 \right)\right)
\bigg]
\bigg\}\,,
\\
\hat\sigma^b_{qg}&=&
\hat\sigma^{(c+)}_{qg}+\hat\sigma^{(f)}_{qg}\,,\\
\hat\sigma^b_{gq}&=&
\hat\sigma^{(c-)}_{gq}+\hat\sigma^{(f)}_{gq}\,,\\
\hat\sigma^b_{q\bar q}&=& \hat\sigma^{(f)}_{q\bar q}\,,\label{eq:sBqq}
\eeeq
where the expressions for ${\cal R}^{(2)}$, ${\cal I}^{(2)}$, and ${\cal V}^{(2)}$ can be found in Ref.~\cite{deFlorian:2013uza}. The factor $V_\text{eff}$ accounts for the effective vertex between gluons and the (on-shell) Higgs boson appearing in $\hat \sigma^b$, and is given by 
\beq
V_\text{eff} = c_t\, F_\triangle (Q/2)+ \f{2}{3}\, 12\,c_g, \label{eq:Veff}
\eeq
where the form factor $F_\triangle$ is evaluated at half the invariant mass of the produced Higgs pair ($Q/2$). 
At variance with all other contributions, where $F_\triangle$ arises from the triangle diagram originated (at LO) from on-shell gluons and, therefore,  is evaluated at the  off-shell Higgs (invariant) mass $Q$, in this effective vertex the outgoing Higgs is on-shell while the exchanged gluon is off-shell. As a consequence, evaluating this vertex either at the fixed scale $M_h$ or the invariant mass $Q$ is not fully satisfactory, and a dynamical scale appears as a more sensible choice. We use the same scale as the one chosen for the renormalization and factorization scales, $Q/2$, which also takes the value $M_h$ at the production threshold.

The expressions for $\sigma^{(c+)}_{ij}$, $\sigma^{(c-)}_{ij}$ and  $\sigma^{(f)}_{ij}$ are given in Appendix \ref{Ap:SigmaB}, and correspond to the renormalized real emission corrections defined in Ref.~\cite{deFlorian:2013jea}, with a modification to properly account for the $V_\text{eff}$ contribution.

Replacing these expressions into Eq.~\eqref{eq:sigma_exp} provides the result for the differential cross section for Higgs pair production in the EFT as a function of the invariant mass of the pair, including the radiative corrections up to NNLO in QCD. 

A comment is in order regarding the rescaling of the HTL result with the exact LO result. In previous work~\cite{deFlorian:2013jea}, the final result has been reweighted with the quotient between the Born cross sections, schematically
\beq \label{eq:HTLimproved}
{\rm d}\sigma = \frac{{\rm d}\sigma_{LO}^\text{Exact}}{{\rm d}\sigma_{LO}^\text{HTL}} {\rm d}\sigma^\text{HTL}.
\eeq
However, this method can fail if at some point of the phase space the Born cross section in the HTL vanishes and the exact does not.
This is not a problem in the SM, for which the LO partonic cross section in the HTL only vanishes at the production threshold, but it is in the EFT, where particular combinations of the coefficients $c_i$ can produce vanishing cross sections in the HTL for certain values of $Q>2M_h$, while they remain different from zero for the exact result (as seen in section 3.3 of Ref.~\cite{Borowka:2016ypz}).
In order to surpass this issue, in this work we directly rescale the individual vertices that appear at the amplitude level.
To be precise, in those matrix elements for which the QCD corrections factorize from the Born cross section, the full LO cross section is introduced by using the exact expression for $C_{LO}$. This is the case for the amplitudes that contribute to $\sigma^{(a)}$.
In the terms arising from the new topology that appears at NLO, the reweighting of the two corresponding gluon-Higgs effective vertices is taken into account by the factor $V_\text{eff}$ defined in Eq.~\eqref{eq:Veff}.
This factor is introduced in the terms extracted from Ref.~\cite{deFlorian:2013uza} either as its square modulus, or as the real and imaginary parts of $C_{LO}^* \,V^2_\text{eff}$ in the interference terms. In particular, in the real emission contributions presented in Appendix~\ref{Ap:SigmaB}, it is the real part of $C_{LO}^* \,V^2_\text{eff}$ that is taken into account.

As mentioned before, this prescription for rescaling the HTL result is different from the one used in Ref.~\cite{deFlorian:2013jea}, and it allows to generalize it to BSM scenarios in which the HTL Born cross section can vanish for a certain invariant mass of the Higgs boson pair.
The difference in the inclusive cross section at NNLO between these two prescriptions is  less than $0.4\%$ (compared to the scale uncertainties of the order of $8\%$-$10\%$). In Ref.~\cite{Grober:2015cwa} a similar prescription was tacitly used, but the effective vertex for the new topology amplitudes $V_\text{eff}$ was introduced in the HTL, as $\frac{2}{3} (c_t + 12 c_g)$. The difference between the rescaling used in Ref.~\cite{Grober:2015cwa} and the one presented herein, results in a slightly stronger dependence of the $K$-factor on the anomalous couplings $c_i$ when using the latter, as we will see in section~\ref{sec:K}.

 \section{Phenomenology}
 \label{sec:pheno}
 
We present here the numerical predictions for the Higgs boson pair production cross section at the LHC, based on the results presented in the previous section. 
For parton densities and strong coupling constant scaling we used the PDF4LHC15 distribution~\cite{Butterworth:2015oua,Dulat:2015mca,Harland-Lang:2014zoa,Ball:2014uwa,Carrazza:2015hva,Watt:2012tq} interpolated with the LHAPDF package~\cite{Buckley:2014ana}. The integration was performed using the CUBA implementation of the VEGAS algorithm~\cite{Hahn:2004fe}. We fixed the collider c.m. energy to $\sqrt{s} = 14$~TeV, the mass of the Higgs and of the top quark were set to the values $M_h = 125$ GeV and $M_t = 172.5$ GeV respectively, and the Higgs width $\Gamma_h = 4.07$ MeV. The factorization and renormalization scales are set to $\mu_R = \mu_F = Q/2$ where $Q$ is the invariant mass of the  Higgs boson pair. 

The anomalous couplings $c_3$ and $c_t$ belong to operators already present in the SM. In fact, $c_3$ modifies the Higgs boson self-coupling and can take values ranging from $c_3=-10$ to $10$~\cite{Azatov:2015oxa}, and $c_t$ modifies the top Yukawa coupling, with allowed values in the interval $0.65\leq c_t \leq 1.15$~\cite{CMS:2014ega}. The normalization is such that the SM corresponds to $c_3 = c_t = 1$, with all the other new couplings set to zero. The rest of the couplings are not present in the SM and only arise due to BSM effects. The parameter $c_{tt}$ corresponds to a new contact interaction between a top-anti-top pair and two Higgs bosons, and is varied from $-1.5$ to $1.5$~\cite{Azatov:2015oxa}. New contact interactions between gluons and one and two Higgs bosons are parametrized through the $c_g$ and $c_{gg}$ couplings, respectively, and both were varied in the range $[-0.15,\,0.15]$~\cite{Azatov:2015oxa}. This interval was chosen mostly for illustrative purposes, despite the fact that the current experimental limit obtained for $c_g$ under certain assumptions is smaller~\cite{Azatov:2015oxa,CMS:2014ega}.

A comment on the validity of the result is in order. The Lagrangian used here is a low energy expansion in powers of $\Lambda$, so the new couplings $c_i$ are expected to show deviations from the SM of order $(v/\Lambda)^2$. With this into account, we can notice that only interferences between the BSM and the SM are of order $\Lambda^{-2}$, whilst quadratic terms in $c_i$, together with the interference between the SM and dimension 8 operators (not considered here), are of order $\Lambda^{-4}$. In principle, one should expand the result linearly in the anomalous couplings and constrain their values to the region where this linear approximation is valid.

\begin{figure}
\begin{center}
\includegraphics[width=.6\columnwidth]{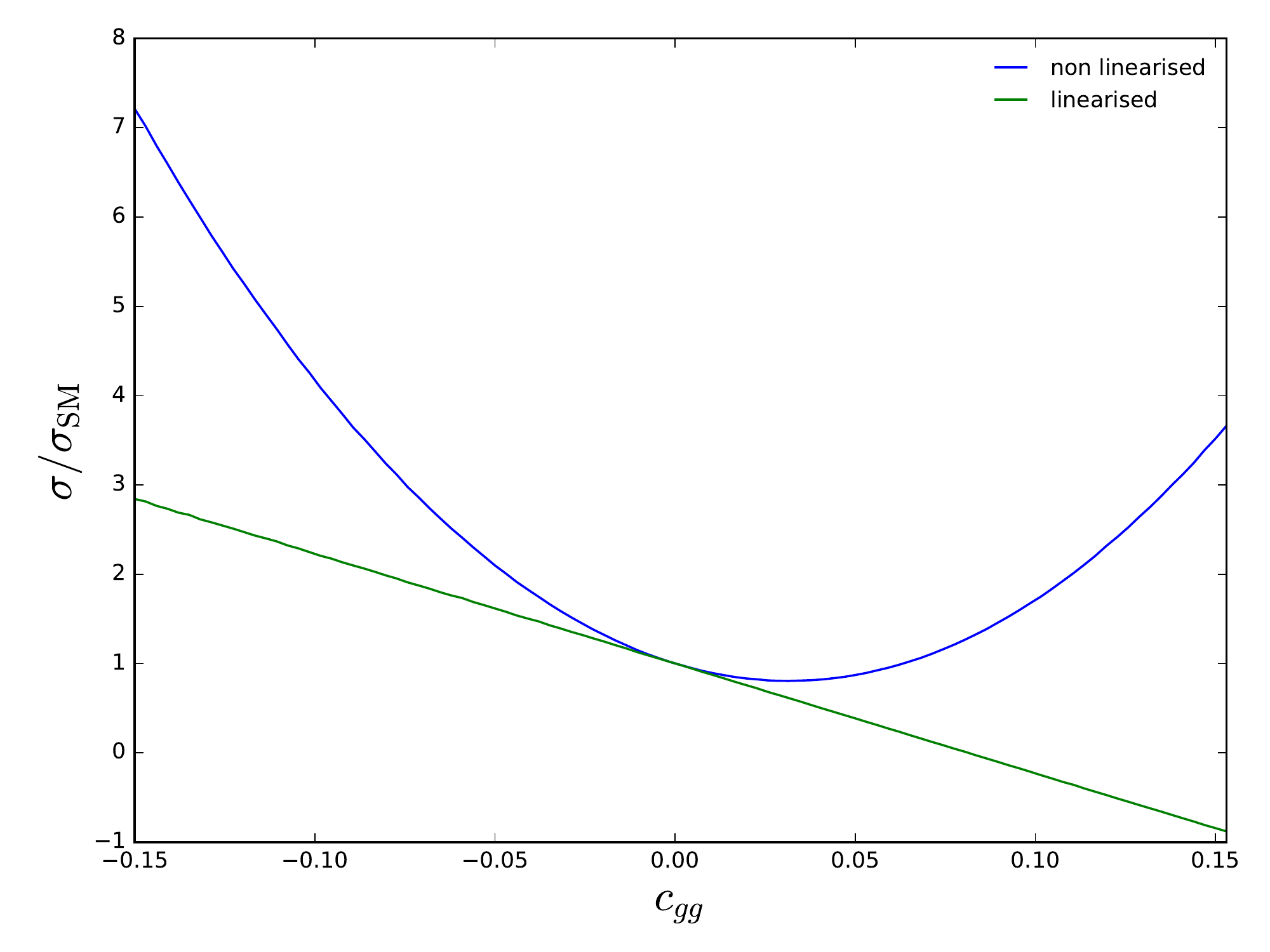}
\end{center}
\caption{\label{fig:Validity}\small
Total cross section for Higgs boson pair production at LO as a function of $c_{gg}$, both for the linearised and non-linearised cases.
}
\end{figure}

 In this paper we vary the coefficients $c_i$ in the whole range so far experimentally allowed, far beyond the region where the linear approximation is valid. This is illustrated in Figure~\ref{fig:Validity}, where we show the LO total cross section for Higgs pair production as a function of $c_{gg}$, both for the linearised and non-linearised cases.
It is clear that we are exploring regions which are far beyond the linear regime; in particular we can observe that the linear approximation vanishes at $c_{gg} \approx 0.08$ and continues to negative values. Nevertheless, in the absence of results for operators up to dimension 8, it has been shown that in some cases it is better to keep the non linear terms~\cite{Azatov:2016sqh}. In particular, in scenarios where the dimension 8 operators do not interfere with the SM (e.g. due to different total helicity) or where the dimension 6 couplings are enhanced (e.g. strongly coupled theories) the current analysis is justified.

\subsection{$K$-factors}
\label{sec:K}
We compute the $K$-factors, as the ratios between the NNLO and LO cross sections. It is interesting to see if the change introduced by the new couplings factorizes from the radiative corrections or if, on the contrary, there is a significant dependence of the $K$-factors on the anomalous couplings. To compare dependencies of the $K$-factor as a function of the different anomalous couplings, we parametrize their departure from the SM with the variable $\xi$. In this computation only one anomalous coupling at a time is left free, and the rest are set to their SM values ($\xi = 0$). The coefficients $c_i$ are parametrized as follow (note that for illustrative purposes, $c_3$ is varied from $-9$ to $11$),
\beeq
c_3 &=& 1 + 10 \; \xi \,, \label{eq:K_paramet_start}\\
c_t &=& 1 + 0.35 \;\xi \,, \\
c_{tt} &=&  1.5\; \xi \,, \\
c_g &=& 0.15 \;\xi \,, \\
c_{gg} &=& 0.15\; \xi \,. \label{eq:K_paramet_end}
\eeeq

\begin{figure}
\begin{center}
\includegraphics[width=.6\columnwidth]{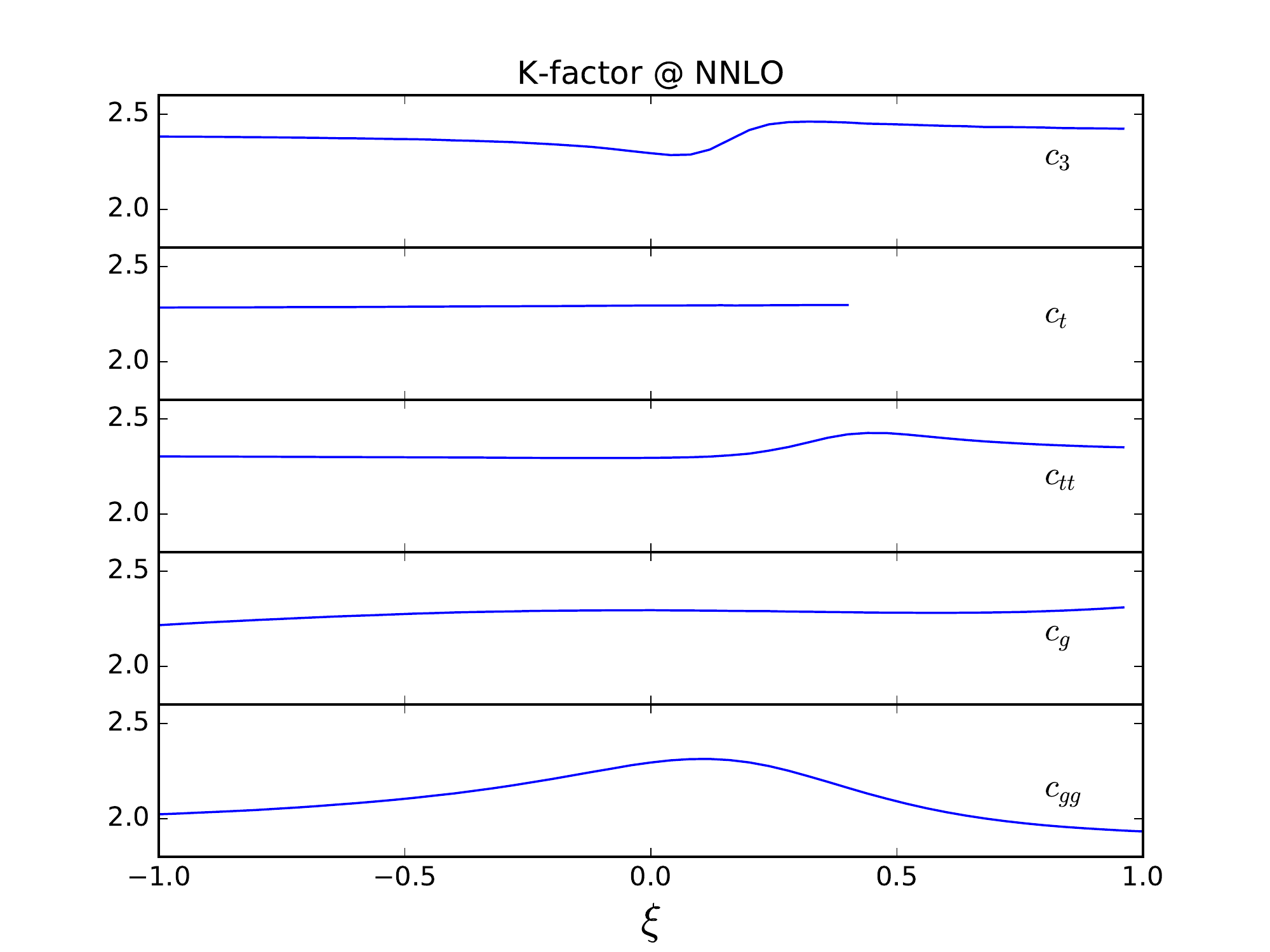}
\end{center}
\caption{\label{fig:kFactors}\small
$K$-factor for total production of Higgs boson pairs as a function of each anomalous coupling, whilst leaving the others set to their SM value. The SM corresponds to $\xi=0$ in all cases, and the parametrization in terms of $\xi$ is given by Eqs.~\eqref{eq:K_paramet_start}~--~\eqref{eq:K_paramet_end}.}
\end{figure}

We can observe from Figure~\ref{fig:kFactors} that the dependence of the $K$-factor on each anomalous coupling, aside for a small bump in some cases, is rather flat. The origin of the bump can be understood on a simple basis: the dependence of the cross section (both at LO as NNLO) on the anomalous couplings is shaped as a parabola with a minimum near the SM. The $K$-factor is then a quotient of two parabola-like functions, thus has a single extreme (either a maximum or a minimum) in $\xi_0$ if the parabolas share a minimum in $\xi_0$, or a maximum next to a minimum if the minima of the parabolas are separated from one another (as is the case of $c_3$). From this kind of analysis, we see that the radiative corrections change the position of the minimnum of the cross section as a function of $c_3$. In the case of $c_t$, the minimum lies outside the allowed values for the anomalous coupling, resulting in a rather flat dependence of the $K$-factor.

The maximum deviation of the $K$-factor from the SM case, when varying one anomalous coupling at a time, is achieved by $c_{gg}$ with a departure of 
\beq
\Delta K^{c_{gg}} = \frac{\text{max}  \left| K(c_{gg}) - K_{SM} \right|}{K_{SM}} \approx 15.8 \% \qquad \text{at } c_{gg} = 0.15\,.
\eeq
The rest of the parameters show the following  maximum departure:

\beeq
\Delta K^{c_{3}}  &\approx 7.2\%  \qquad &\text{at } c_{3}\,=\;4.20 \,,\\
\Delta K^{c_{tt}} &\approx 5.7\%  \qquad &\text{at } c_{tt} =\;0.66 \,,\\
\Delta K^{c_{g}}  &\approx 3.4\%  \qquad &\text{at } c_{g}\,= -0.15 \,,\\
\Delta K^{c_{t}}  &\approx 0.5\%  \qquad &\text{at } c_{t}\;=\;0.65 \,.
\eeeq

Nevertheless, when we allow the five anomalous couplings to vary at the same time, larger deviations from the SM $K$-factor can be found. When sampling the five dimensional parameter space\footnote{Clustering algorithms that are sensitive to the kinematics of the process have been derived in~\cite{Carvalho:2015ttv} that could improve the choice of the benchmark points when sampling high-dimensional parameter spaces.}, the maximum deviation observed is
\beq
\Delta K^{\text{max}} \approx 84\%
\eeq
 at the point of parameter space $c_3=7.0,\; c_t= 1.15,\;c_{tt}= 0.1,\;c_g= -0.09,\;c_{gg}= 0.02$. At this point, the $K$-factor is as high as $4.07$, and if we modify the value of $c_g$ to $c_g = -0.11$ we get a value as low as $0.80$ ($36\%$ of the SM value). This shows a strong dependence on $c_g$ on this region of parameter space, with a similar shape as the dependence shown for $c_3$ in Figure~\ref{fig:kFactors}. The reason is also clear: at the maximum $K$-factor, the LO cross section has a minimum of $2.46$ fb ($12.5\%$ of the SM value), whilst at the minimum of the $K$-factor (lowering $c_g$) it is the NNLO cross section that gets minimized to $4.84$ fb ($10.7\%$ of the SM value). Because the cross section is near a minimum, any variation in the radiative corrections amounts for a significant relative change, resulting in this  behaviour of the $K$-factor. It is also worth noting that in this region of parameter space where the $K$-factor reaches a minimum of $0.76$, the radiative corrections are negative, thus resulting in a decrease of the total cross section with respect to the LO one. Also, the NLO $K$-factor at that point is $1.01$, still greater than one, which means that the negative corrections that decrease the cross section below its LO value are a purely NNLO effect. The change of sign of the NNLO corrections while varying the value $c_g$ from $-0.09$ to $-0.11$ at this point of parameter space is driven by the sign of the contributions proportional to ${\rm Re}\left(A^* C_{LO}\right)$ in $\hat\sigma^a$ (Eq.~\eqref{eq:sigmaA}), which happen to be dominant in this particular region.

From this analysis we conclude that, while the $K$-factor can be approximated by a constant (up to a $16\%$ variation) when moving one coupling at a time away from its SM value, that no longer holds true if we allow a general deviation from the SM. When considering situations in which the cross section is small, the shape of the $K$-factor plays an important role and should be taken into account. 
Of course, the dependence of the $K$-factor on the EFT parameters presented here may be subject to modifications once the full top-mass effects are included at NLO.

\subsection{Degeneracy of the parameters}
\label{sec:degenerate}

\begin{figure}
\begin{center}
\begin{tabular}{c c}
\begin{tabular}{c}
\includegraphics[width=0.5 \columnwidth]{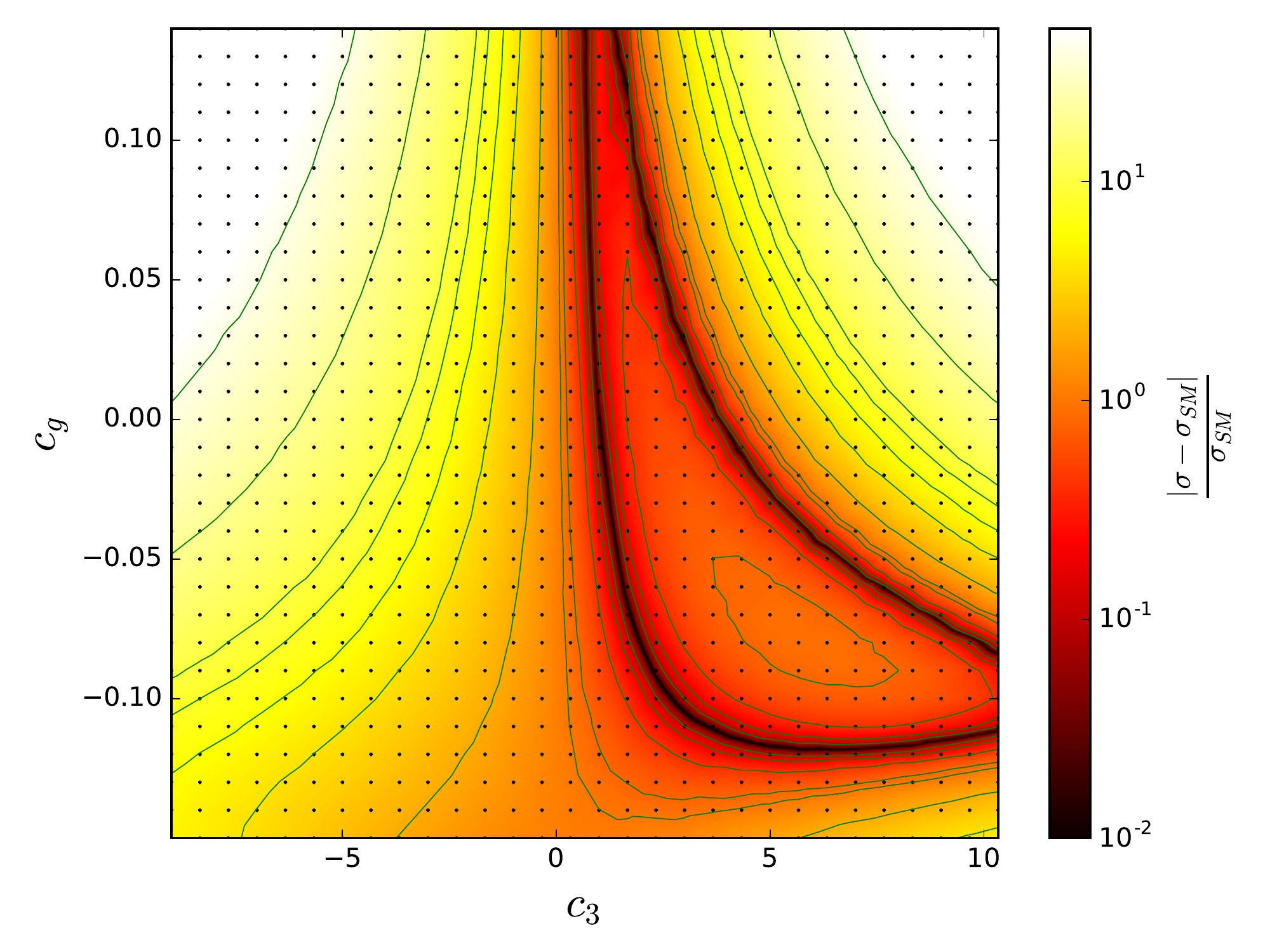}
\\
\hspace*{-0.5cm}
\includegraphics[width=0.5 \columnwidth]{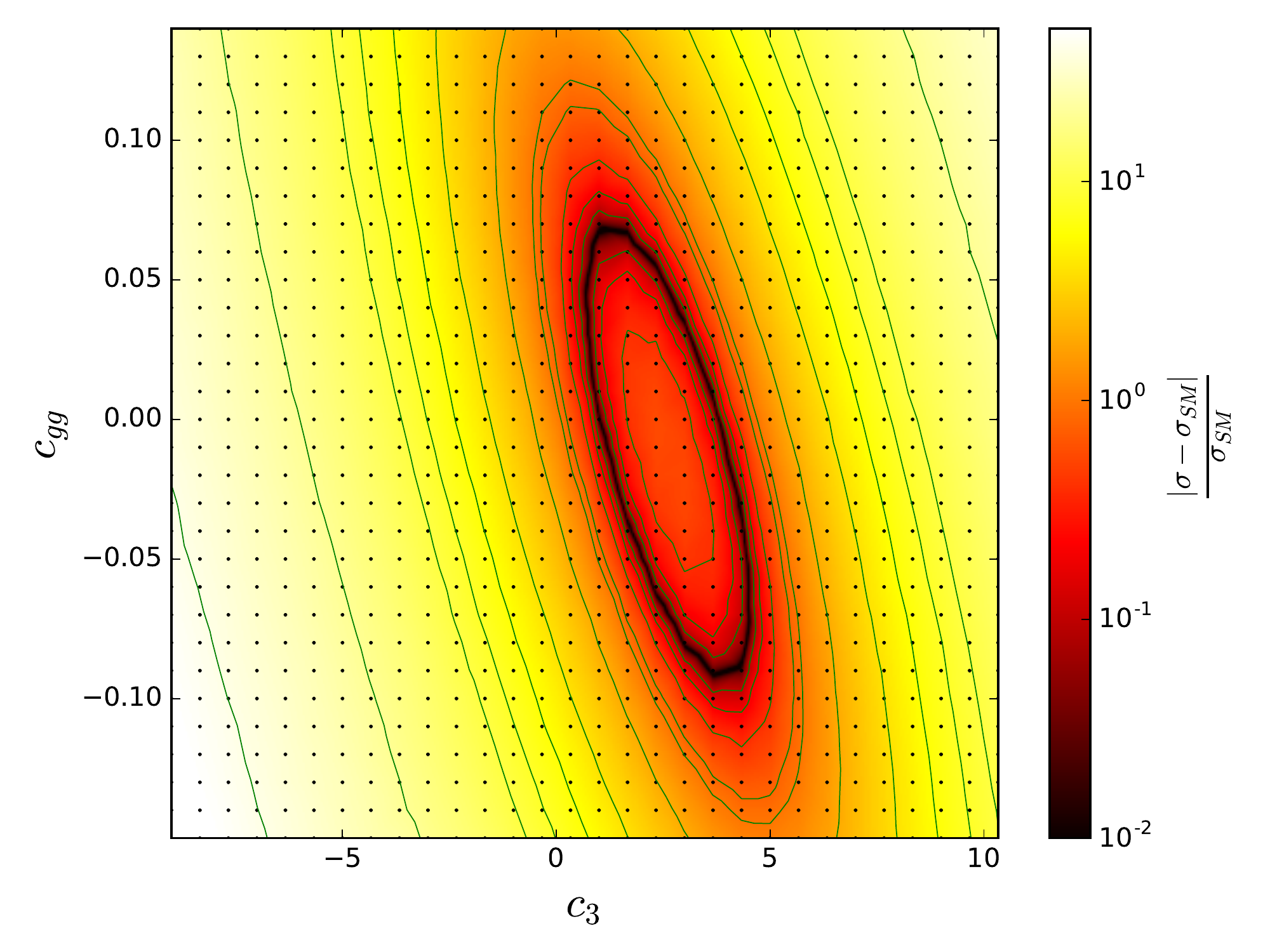}
\end{tabular}
&
\begin{tabular}{c}
\includegraphics[width=0.5 \columnwidth]{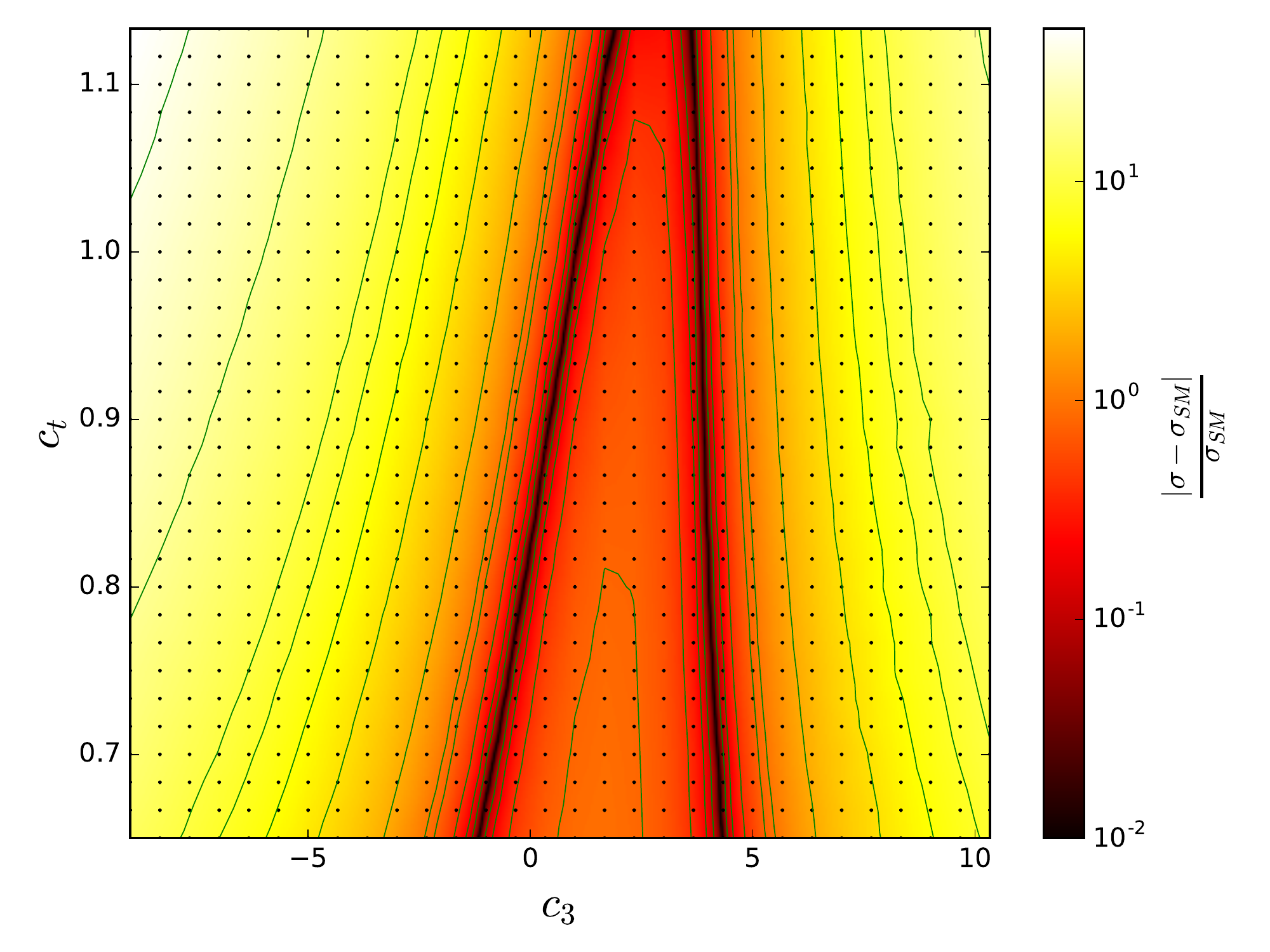}
\\
\hspace*{-0.5cm}
\includegraphics[width=0.5 \columnwidth]{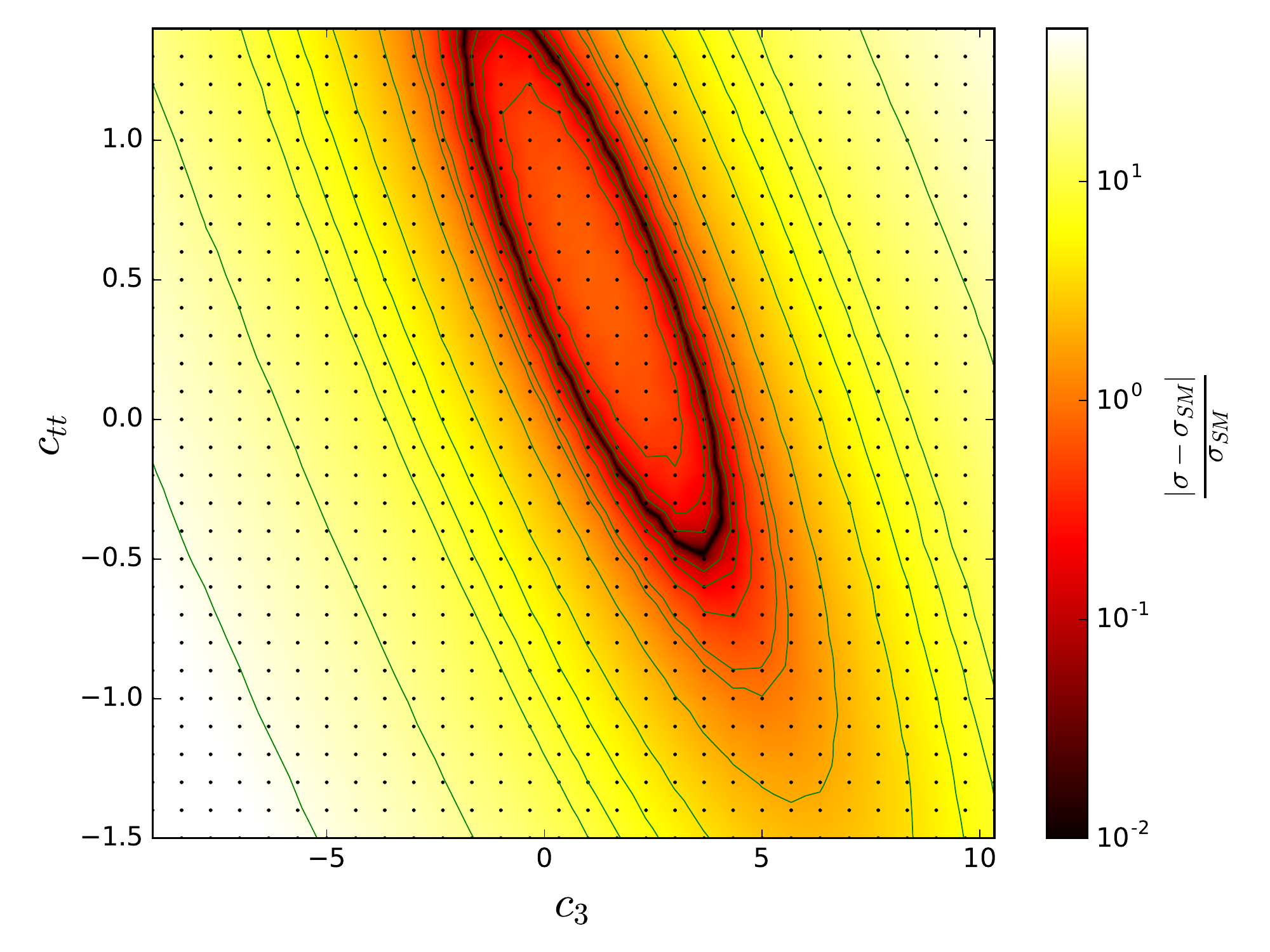}
\end{tabular}
\end{tabular}
\end{center}
\caption{\label{fig:degeneracy}\small
Heatmaps showing the total cross section in a coloured logarithmic scale as a function of pairs of anomalous couplings. The degenerate directions are drawn in green lines, in particular the black region of the heatmaps shows some combination of parameters degenerate with the SM. The points mark the grid in which the cross section was computed, and a cubic interpolation was done elsewhere for illustrative purposes.
}
\end{figure}

One important question that arises when considering several parameters is that of their degeneracy. 
The  partonic cross section at LO in the HTL can be written as
\beq
\frac{{\rm d} \sigma_{LO}}{{\rm d}Q} = \frac{{\rm d}\sigma_{SM}}{{\rm d}Q}\, \frac{\left| \left[ c_3 (c_t + 12\, c_{g})\right]C_{\triangle} + \left[-c_t^2 + c_{tt} + 12\,c_{gg}\right] \right|^2}{\left| C_{\triangle} - 1 \right|^2},
\label{eq:degeneracy}
\eeq
where $C_{\triangle}$ is defined in Eq.~\eqref{eq:Ctriangulo} and does not depend on the anomalous couplings. We see that the total cross section depends only on two linear combinations of the couplings, the ones inside the squared brackets. 

In  order to see the structure of the degeneracy at NNLO, we present in Figure~\ref{fig:degeneracy} heatmaps of the relative deviation of the total Higgs pair production cross section from its SM value in four different two-dimensional slices of the parameter space. These slices correspond to the variation of the anomalous Higgs boson self-coupling together with the other four anomalous couplings separately.

We can see in Figure~\ref{fig:degeneracy} that the structure of the degeneracy presented at LO in the HTL is preserved after including radiative corrections and reweighting by the exact Born cross section. In the two lower plots we see an elliptic pattern of degeneracies, which is related to the fact that the two couplings varied ($c_3$ and the couplings of gluons and top quarks to a pair of Higgs bosons) modify two different topologies of diagrams (triangle and box-like) and thus enter in two different terms in the amplitude. If we expand the square in Eq.~\eqref{eq:degeneracy} setting all other couplings to their SM values, we see that the expression is quadratic in the previously mentioned couplings, leading to an elliptic pattern of degeneracies. In the two upper plots, the two couplings varied modify the same diagram and enter in the final expression multiplicatively, resulting in a deformed pattern with respect to the two lower plots. It is easy, for example, to recognize in the first plot the family of parameters degenerated with the SM arising from the relation $c_3 (1 + 12 \,c_g)= 1$. 

A consequence of the present degeneracies on the anomalous couplings is that, even in the case of a measurement for the total cross section compatible with the SM prediction, it would be possible to accommodate significant departures from the SM couplings (the dark bands on the heatmaps of Figure~\ref{fig:degeneracy}) without affecting the corresponding theoretical prediction. This means that the total cross section is not enough to distinguish between different scenarios and more observables are needed, e.g. differential distributions (see Refs.~\cite{Grazzini:2016paz,DiVita:2017eyz}).

\subsection{Invariant mass distributions}
\label{sec:invariant}

As we could see from section~\ref{sec:degenerate}, the total cross section is not enough to discriminate between the effects of the different EFT parameters, and therefore differential distributions are needed. Our calculation allows us to compute the invariant mass distribution of the produced Higgs boson pair, thus providing a tool for breaking these degeneracies. In fact, because of the scalar character of the Higgs boson, there is no reason to expect strong angular dependencies and most of the information of the process can be extracted from the invariant mass  and transverse-momentum distributions (see Ref.~\cite{Grazzini:2016paz}).

In the SM, an almost exact destructive interference between the box and triangle diagrams occurs at the production threshold~\cite{Dawson:2015oha}, resulting in an overall small cross section. At tree-level in the SM, because the triangle and box amplitudes are independently gauge invariant, we can write the cross section in terms of their contributions $\mathcal{M}_{\triangle}$ and $\mathcal{M}_{\square}$ as
\beq\label{eq:Interference}
{\rm d}\sigma_{SM} \propto \left| \mathcal{M}_{\triangle} \right|^2 + \left|\mathcal{M}_{\square}\right|^2 + 2 {\rm Re}\left(\mathcal{M}_{\triangle}^*\mathcal{M}_{\square}\right)
\eeq
When changing the Higgs boson triple self-coupling, the cancellation is still destructive~\cite{Dicus:2015yva} but the balance between the triangle (which is affected by the self-coupling) and box (independent of the self-coupling) contributions changes, resulting in a fast increase in the cross section~\cite{Dawson:2015oha}. Also, when new (scalar or vector-like fermions) particles that couple to the Higgs boson are taken into account inside the loop (which corresponds to modifications of  $c_g$ and $c_{gg}$ in the EFT), the cancellation between contributions also breaks down and the cross section grows~\cite{Dawson:2015oha}. This makes the threshold region very sensitive to BSM physics.

\begin{figure}
\begin{center}
\begin{tabular}{c c}
 \includegraphics[width=.45\columnwidth]{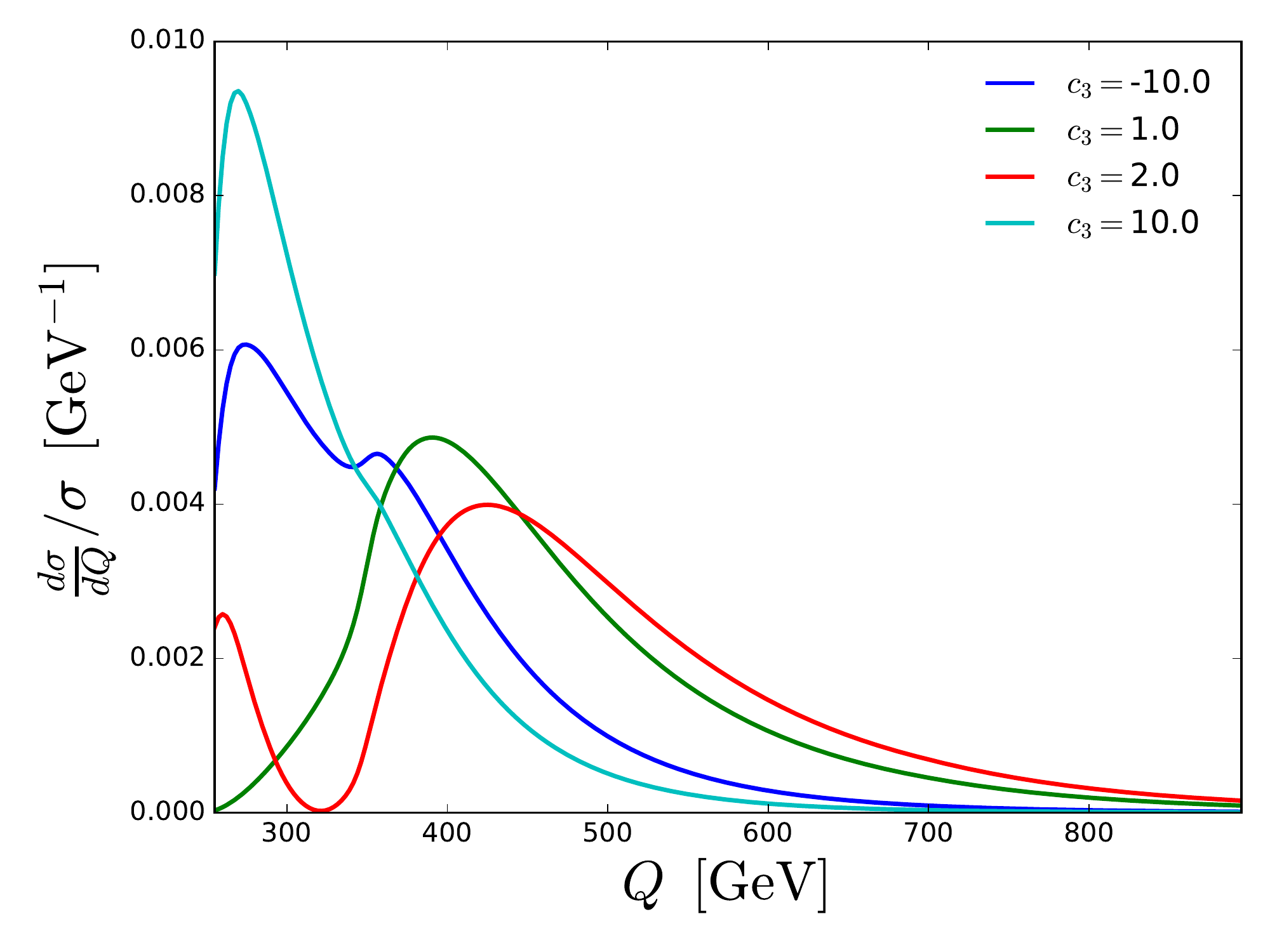}
& \includegraphics[width=.45\columnwidth]{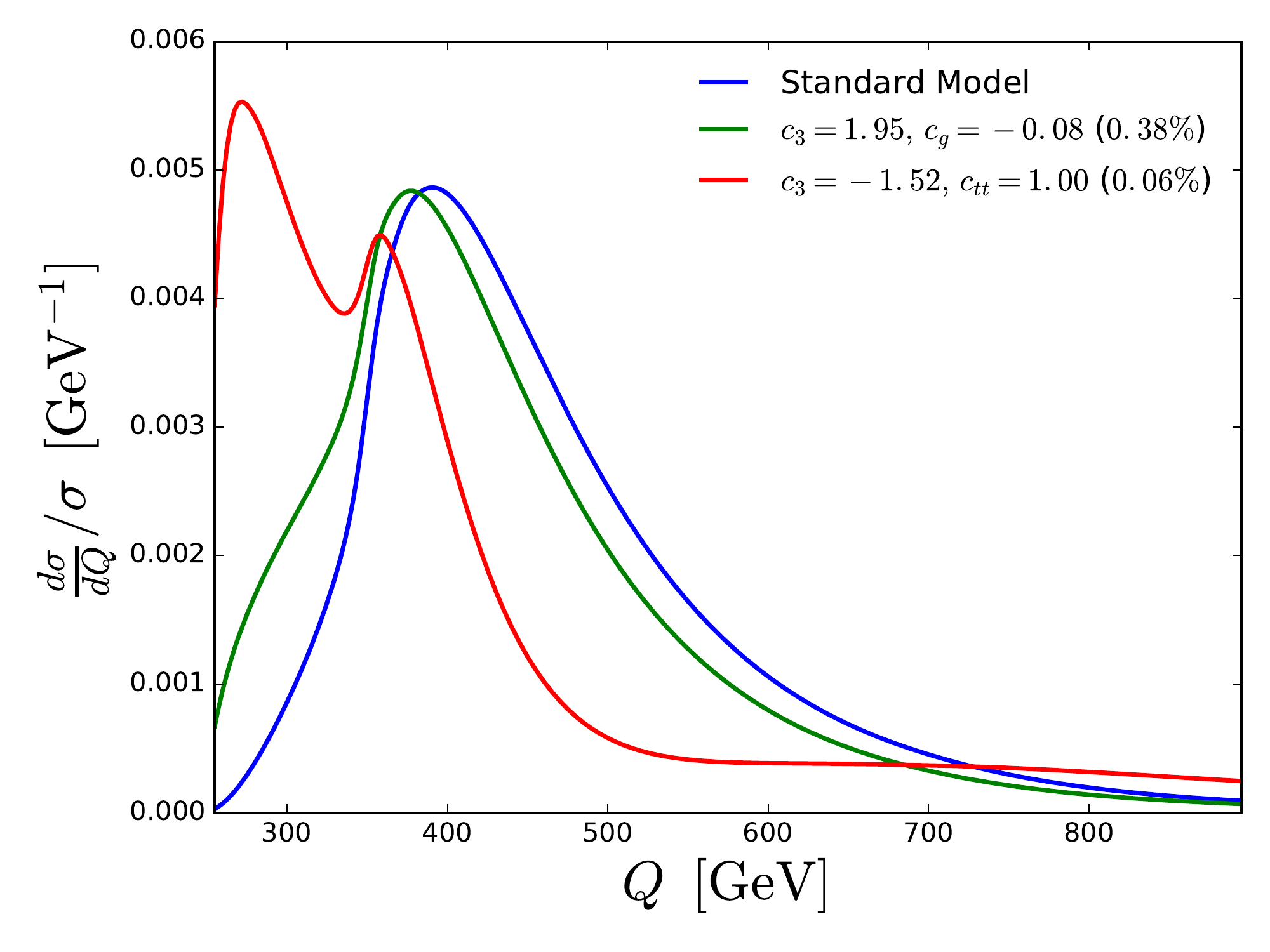}\\
(a) & (b)
\end{tabular}
\end{center}
\caption{\label{fig:InvMass}\small
Invariant mass distribution of the produced Higgs boson pair plotted for (a) different values of its self-coupling, and (b) different combinations of anomalous couplings that are degenerate with the SM. The relative deviation from the SM of the total cross section in (b) is specified between brackets on the label.}
\end{figure}

In Figure~\ref{fig:InvMass}(a) we show the invariant mass distributions for different values of the Higgs boson anomalous self-coupling ($c_3$). As mentioned before, varying the self-coupling changes the balance between amplitudes in Eq.~\eqref{eq:Interference}, resulting in more involved escenarios. 
When the self-coupling runs to large values, either positive or negative, the triangle amplitude dominates and we observe a boost of the cross section at threshold due to the Higgs boson propagator in the triangle contribution.

In Figure~\ref{fig:InvMass}(b) we show combinations of anomalous couplings that render inclusive cross sections similar to the one of the SM (their relative deviations from the SM are shown between brackets on the label). They correspond to points of parameter space inside the black bands in Figure~\ref{fig:degeneracy}. Nevertheless, we can identify rather different behaviours for each of their invariant mass distributions. 
For instance for the green curve in Figure~\ref{fig:InvMass}(b), when choosing $c_{g} = -0.8 \approx -\frac{1}{12}$, there is an almost exact cancelation between this term and the one proportional to $F_\triangle$ at LO in the HTL (see Eq.~\eqref{eq:CLO}), and thus $\mathcal{M}_{\triangle} \approx 0$. The resulting invariant mass distribution is given mostly by the box contribution ($\left|\mathcal{M}_{\square}(Q)\right|^2$). 
In the other case, for the red curve, the value $c_{tt} = 1$ sets $\mathcal{M}_{\square}$ to zero in the HTL, and this results in a distribution given primary by the triangle contribution ( $\left|\mathcal{M}_{\triangle}(Q)\right|^2$), which shows the two peaks that arise from the pole on the Higgs propagator, the first one, and from the maximum of $\left|F_\triangle(Q)\right|$, the second one.

The substantial differences that we observe for these particular points in the parameter space in the shape of the invariant mass distributions shown in Figure~\ref{fig:InvMass} illustrate the fact that this observable can definitely help to disentangle the contributions from different operators which otherwise would be degenerated.

\section{Conclusions}
\label{sec:conc}

Although BSM physics might not be available for direct detection through resonances in the near future, physics at a higher energy scale can be accessed through precision measurements of the Higgs boson couplings. The triple Higgs boson coupling is of particular interest because it provides insight into the electroweak symmetry breaking mechanism, and Higgs boson pair production results in a sensitive channel for these studies. The EFT approach supplies a model independent way of addressing these issues, by the addition of higher dimensional effective operators. In order to be consistent one must include all operators that modify the Higgs boson couplings relevant for the process, and for Higgs boson pair production through gluon fusion there are five relevant dimension 6 operators. Then, the Wilson coefficients of these operators can be fitted to experimental data.

In this work we have computed the cross section for Higgs boson pair production, both inclusive and differential in the invariant mass of the produced pair, including all terms up to $\mathcal{O}(\alpha_S^4)$ (NNLO in QCD) and considering all relevant dimension 6 operators. These modify the coupling between the Higgs boson and the top quark, as well as the Higgs triple self-coupling, and they add new contact interactions between the Higgs and the gluon field. The calculation was performed in the HTL and a proper rescaling prescription was used to approximate the effects of the finite top quark mass on the result. Then, a comprehensive study of the phenomenology introduced by the anomalous couplings was carried out, from which we would like to emphasize the following points:

\begin{itemize}

 \item The $K$-factor as a function of the couplings was found to be rather flat, within a $16\%$ deviation from its SM value. The exception is in regions of the anomalous couplings space where the cross section is minimized and radiative corrections turn out to be significant, reaching values for the $K$-factor as high as $4.07$ ($84\%$ higher than the SM) and as low as $0.76$ ($34\%$ of the SM value).

\item The degeneracies of the inclusive cross section with respect to the anomalous couplings were studied, showing that radiative corrections do not substantially alter its shape. Also, it was shown that even in the case of a measurement compatible with the SM inclusive cross section expectation, it is possible to have large deviations from the SM couplings that render the same value.

\item The differential invariant mass distribution of the produced Higgs boson pair was studied, showing that it encodes enough information to disentangle between combinations of anomalous couplings that are degenerate in the inclusive cross section. The reason for this high sensitivity is the destructive interference between triangle and box diagrams, that renders a small cross section near threshold in the SM. This cancellation, and the corresponding large deviation from it in BSM scenarios, provides a powerful tool for the discovery of new heavy physics that couples to the Higgs boson.
\end{itemize}

Some final comments must be held about the limitations of the present calculation, in which we are including the contributions from amplitudes mediated by dimension 6 operators, both from their interference with the SM as well as from their square modulus. The latter is \emph{a priori} expected to be of the same order as the interference between amplitudes originated from dimension 8 operators and the SM. While in some cases (e.g. strongly coupled theories) the dimension 8 operators can be ignored, in general one should either perform a global analysis including them, or restrict the validity of the calculation to deviations from the SM far smaller than the current experimental bounds.
Regarding the dimension 6 operators included (or not) in this analysis, some dimension 6 QCD operators such as the chromomagnetic dipole-moment, despite being constrained by top quark pair production and representing a higher order correction on the top quark Yukawa coupling, could still result in a significant contribution to the Higgs boson pair production cross section. In order to consistently include this operator into the analysis one should consider several other QCD operators that are mixed with it through Renormalization Group flow (see Ref.~\cite{Maltoni:2016yxb}). The inclusion of such operators, as well as higher dimensional ones, is left for future work.

\acknowledgments

The work of D.deF. has been partially supported by Conicet, ANPCyT and the von Humboldt Foundation.  D.deF. wishes to thank the University of Z\"urich  for the hospitality during the completion of this work.
 
\appendix
\section{Real emission corrections}

\label{Ap:SigmaB}
In the following section we present the expressions for the renormalized real emission contributions to the NNLO cross section. Namely $\hat\sigma^{(c+)}_{ij}$, $\hat\sigma^{(c-)}_{ij}$ and $\hat\sigma^{(f)}_{ij}$, that appear in Eqs.~\eqref{eq:sBgg}~--\eqref{eq:sBqq} for the different partonic subprocesses $ij\to HH + X$. 

The expressions for the SM are presented in Ref.~\cite{deFlorian:2013jea} and are calculated in the HTL and then reweighted by the Born cross section following Eq.~\eqref{eq:HTLimproved}. For the contributions under consideration, this implies that the HTL real correction is multiplied by the factor $\frac{\text{Re}(C_{LO})}{|C_{LO}^{HTL}|^2}$. As discussed at the end of section~\ref{sec:calc}, we use a different prescription to avoid the numerically dangerous division by $|C_{LO}^{HTL}|^2$ and directly introduce the exact LO amplitude as $\text{Re}(C_{LO}^* V_\text{eff}^2)$, also taking into account the reweighting of the effective vertex between gluons and the Higgs of Eq.~\eqref{eq:Veff}. The results are the same as the presented in Ref.~\cite{deFlorian:2013jea}, but making the following replacements
\beeq 
\Delta_{LO} &\to& 1\label{eq:replacements-in} \,,\\
{\rm Re}(C_{LO}) &\to& {\rm Re}\left( C_{LO}^* V_\text{eff}^2 \right)\,,\\
\f{\hat{\sigma}_{LO}}{\left|C_{LO}\right|^2} &\to& \int_{t_-}^{t_+}\!\!dt \f{G_F^2\, \as^2}{512(2\pi)^3}\,,\label{eq:replacements-out}
\eeeq
and then using the definition for $C_{LO}$ given in Eq.~\eqref{eq:CLO}.

For completeness, we present the resulting expressions once these replacements are performed. These can be written (for $\mu_R = \mu_F = Q$) as
\beeq
\hat\sigma^{(c+)}_{gg}&=&
\hat\sigma^{(c-)}_{gg}=
\int_{t_-}^{t_+}\!\!dt \f{G_F^2\, \as^2}{512(2\pi)^3}\,
\left(\f{\as}{2\pi}
\right)^2 
8\left[1-(1-x)x\right]^2 \nonumber\\
&\times&
\left[
2\left(
\f{\log(1-x)}{1-x}
\right)_+
-\f{\log x}{1-x}
\right]
\text{Re}\left( C_{LO}^* V_\text{eff}^2 \right)\,,\nonumber
\\
\hat\sigma^{(f)}_{gg}&=&
\int d\cos\theta_1\,d\theta_2\,dy\,
\f{\sqrt{x(x-4M_H^2/Q^2)}}
{1024\,\pi^4}
\left(\f{1}{1-x}
\right)_+\nonumber\\
&\times&\left[
\left(\f{1}{1-y}\right)_+
+\left(\f{1}{1+y}\right)_+
\right]
f_{gg}(x,y,\theta_1,\theta_2)\nonumber\,,
\\
\hat\sigma^{(c+)}_{qg}&=&
\hat\sigma^{(c-)}_{gq}=
\int_{t_-}^{t_+}\!\!dt \f{G_F^2\, \as^2}{512(2\pi)^3}\,
\left(\f{\as}{2\pi}
\right)^2 
\f{16}{9}\,\Big\{\left[1+(1-x)^2\right]\nonumber\\
&\times&
\left[
2\log(1-x)-\log x
\right]
+x^2\Big\}
\,\text{Re}\left( C_{LO}^* V_\text{eff}^2 \right)\,,\nonumber\\
\hat\sigma^{(f)}_{qg}&=&
\int d\cos\theta_1\,d\theta_2\,dy\,
\f{\sqrt{x(x-4M_H^2/Q^2)}}
{512\,\pi^4}
\left(\f{1}{1-y}\right)_+
f_{qg}(x,y,\theta_1,\theta_2)\,,\nonumber\\
\hat\sigma^{(f)}_{gq}&=&
\int d\cos\theta_1\,d\theta_2\,dy\,
\f{\sqrt{x(x-4M_H^2/Q^2)}}
{512\,\pi^4}
\left(\f{1}{1+y}\right)_+
f_{gq}(x,y,\theta_1,\theta_2)\nonumber\,,
\\
\hat\sigma^{(f)}_{q\bar q}&=&
\int d\cos\theta_1\,d\theta_2\,dy\,
\f{\sqrt{x(x-4M_H^2/Q^2)}}
{512\,\pi^4}\,
f_{q\bar q}(x,y,\theta_1,\theta_2)\,,
\eeeq
where the integration variable $t$ is defined in Eq.~\eqref{eq:t} with limits $t_{\pm} = t\,(\cos\theta_1 = \pm 1)$, $G_F$ is the Fermi coupling, and the plus distributions are defined as 
\beeq
&&\int_0^1 dx\, G_+(x)\, f(x)
=\int_0^1 dx\,
G(x)
\left[f(x)-f(1)\right]\,,\\
&&\int_{-1}^1 dy\, f(y)
\left(
\f{1}{1\pm y}
\right)_+
=\int_{-1}^1 dy\,
\f{f(y)-f(\mp 1)}{1\pm y}\,.
\eeeq 

The functions $f_{ij}(x,y,\theta_1, \theta_2)$ are defined as
\beeq
f_{gg}(x,y,\theta_1,\theta_2) &=&
\frac{\as^4 G_F^2 \text{Re}\left( C_{LO}^* V_\text{eff}^2 \right)}{576 \pi ^2 s}
s (1-x)^2 (1-y^2)\nonumber\\
&\times&
\big[F(s,q_1,q_2,t_k,u_k)+F(s,\hat{q}_1,\hat{q}_2,t_k,u_k)+F(s,q_2,q_1,u_k,t_k)\nonumber\\
&+&F(s,\hat{q}_2,\hat{q}_1,u_k,t_k)
+F(t_k,q_1,w_2,s,u_k)+F(t_k,\hat{q}_1,w_1,s,u_k)\big]\;,\nonumber\\
f_{qg}(x,y,\theta_1,\theta_2) &=&
\frac{\as^4 G_F^2 \text{Re}\left( C_{LO}^* V_\text{eff}^2 \right)}{648 \pi ^2 s}
s (1-x) (1-y)\nonumber\\
&\times&
\left[
h(s,q_1,q_2,t_k,u_k)+h(s,\hat{q}_1,\hat{q}_2,t_k,u_k)
\right]\;,\nonumber\\
f_{gq}(x,y,\theta_1,\theta_2) &=&
\frac{\as^4 G_F^2 \text{Re}\left( C_{LO}^* V_\text{eff}^2 \right)}{648 \pi ^2 s}
s (1-x) (1+y)\nonumber\\
&\times&
\left[
h(s,\hat q_2,\hat q_1,u_k,t_k)+h(s,q_2,q_1,u_k,t_k)
\right]\;,\nonumber\\
f_{q\bar q}(x,y,\theta_1,\theta_2) &=&
-\frac{\as^4 G_F^2 \text{Re}\left( C_{LO}^* V_\text{eff}^2 \right)}{243 \pi ^2 s}
s (1-x) \nonumber\\
&\times&
\left[
\, h(t_k,q_1,w_2,s,u_k)+\, h(t_k,\hat q_1,w_1,s,u_k)
\right]\;,
\eeeq
and the function $F$ is defined as follows
\beq
F(s,q_1,q_2,t_k,u_k)=f_1(s,q_1,q_2,t_k,u_k)
+f_2(s,q_1,q_2,t_k,u_k)\;.
\eeq
The invariants used \cite{Frixione:1993yp} are defined in terms of $Q^2$, $x$, $y$, $\theta_1$ and $\theta_2$ as
\beeq
s &=& Q^2\;,\nonumber\\
t_k &=& -\tfrac{1}{2}s(1-x)(1-y)\;,\nonumber\\
\!u_k &=& -\tfrac{1}{2}s(1-x)(1+y)\;,\nonumber\\
q_1 &=& M_H^2-\tfrac{1}{2}(s+t_k)(1-\beta_x \cos\theta_1)\;,\nonumber\\
q_2 &=& M_H^2-\tfrac{1}{2}(s+u_k)(1+\beta_x \cos\theta_2\sin\theta_1\sin\psi
+\beta_x\cos\theta_1\cos\psi)\;,\nonumber\\
\;\hat{q}_1 &=& (p_1-k_2)^2=2M_H^2-s-t_k-q_1\;,\nonumber\\
\;\hat{q}_2 &=& (p_2-k_1)^2=2M_H^2-s-u_k-q_2\;,\nonumber\\
\!w_1 &=& (k+k_1)^2=M_H^2-q_1+q_2-t_k\;,\nonumber\\
\!w_2 &=& (k+k_2)^2=M_H^2+q_1-q_2-u_k\;,
\eeeq
where the coefficients $\beta_x$ and $\psi$ defined as
\beeq
\beta_x&=&\sqrt{1-\f{4M_H^2}{x\,s}}\,,\\
\cos\psi&=&1-\frac{8 x}{(1+x)^2-(1-x)^2 y^2}\,.\nonumber
\eeeq

Finally, the expressions for the functions $f_1$, $f_2$ and $h$ that complete the presentation of the real emission contributions are the following
\beeq
&&f_1(s,q_1,q_2,t_k,u_k)=
\f{1}{q_1 s\, t_k (M_H^2 + q_1 - q_2 - u_k) u_k}
\Big[
s t_k (-q_2^2 (2 s-3 t_k) (s+t_k)+q_1 q_2 \nonumber\\&&\;\;
(6 s^2+3 s t_k+2 t_k^2+q_2 (s+t_k))-q_1^2(q_2 s+4 (s^2+s t_k+t_k^2)))+(2 (q_1-q_2) s^2 \nonumber\\&&\;\;
(2 q_1^2-2 q_1 q_2+q_2^2+s^2)+s(-q_1^2 s+q_2 (-3 q_2^2+2 q_2 s-8 s^2)+q_1 (6 q_2^2+3 q_2 s+14 s^2)) t_k\nonumber\\&&\;\;
+(-8 q_1^2 s-q_2 (q_2^2-3 q_2 s+7 s^2)+q_1 (q_2^2+10 q_2 s+17 s^2)) t_k^2+(-4q_1^2+6 q_1 (q_2+s)\nonumber\\&&\;\;
+q_2 (q_2+4 s)) t_k^3) u_k+(2 s (2 q_1^3-2 q_1^2 (q_2+s)-2q_2 s (q_2+s)+q_1 (q_2+s) (q_2+3 s))\nonumber\\&&\;\;
+(q_1 (q_1-q_2) q_2+(11 q_1-6 q_2)q_2 s+2 (11 q_1-3 q_2) s^2-2 s^3) t_k+(-4 q_1^2+7 q_1 q_2-3 q_2^2\nonumber\\&&\;\;
+23 q_1 s+q_2 s-6 s^2) t_k^2+(6 q_1+2 q_2+s) t_k^3) u_k^2+(-4 s (q_1^2+q_2 s-q_1 (q_2+2s))\nonumber\\&&\;\;
-(3 q_1^2-13 q_1 s+s (7 q_2+4 s)) t_k+(6 q_1-3 q_2-2 s) t_k^2+t_k^3) u_k^3+(q_1-t_k)(4 s+t_k) u_k^4\nonumber\\&&\;\;
-M_H^6 s (t_k+u_k) (t_k+2 u_k)+M_H^4 (s t_k ((-q_1+q_2)s+(q_1+2 q_2-2 s) t_k+3 t_k^2)\nonumber\\&&\;\;
+(2 (q_1-q_2) s^2+3 (2 q_1+q_2) s t_k+(q_1-q_2+9s) t_k^2+5 t_k^3) u_k+(q_1 (6 s+t_k)\nonumber\\&&\;\;
+t_k (-q_2+9 s+6 t_k)) u_k^2+t_k u_k^3)+M_H^2(s t_k ((q_1-q_2) (q_1+q_2-2 s) s+(-q_2 (2 q_1+q_2)\nonumber\\&&\;\;
+(q_1+q_2) s) t_k+2(q_1-3 q_2) t_k^2)-(4 q_1 (q_1-q_2) s^2+s (-3 q_2 (-4 q_1+q_2)+(q_1+3q_2) s\nonumber\\&&\;\;
-2 s^2) t_k+2 (-q_2^2+6 q_2 s-4 s^2+q_1 (q_2+s)) t_k^2+2 (q_1+3 q_2+2s) t_k^3) u_k-(2 s (4 q_1^2\nonumber\\&&\;\;
-2 q_1 q_2+q_2^2+s^2)+(q_1^2-q_2 (q_2-3s)+11 q_1 s) t_k+(3 (q_1+q_2)+8 s) t_k^2+6 t_k^3) u_k^2\nonumber\\&&\;\;
+(-4 s (q_2+s)-6 s t_k-5t_k^2+2 q_1 (2 s+t_k)) u_k^3-(4 s+t_k) u_k^4)
\Big]\;,\\
&&f_2(s,q_1,q_2,t_k,u_k)=
\f{1}{q_2 s\, t_k^2 u_k}
\Big[
s t_k (4 q_2 s^3-s ((M_H^2+3 q_1-4 q_2) (q_1-q_2)\nonumber\\&&\;\;
+(4 M_H^2+q_1-11 q_2) s) t_k-((M_H^2+3q_1-4 q_2) (M_H^2-q_2)+(7 M_H^2+2 q_1\nonumber\\&&\;\;
-11 q_2) s) t_k^2 +4 (-M_H^2+q_2) t_k^3)-(4(q_1-q_2)^2 s^3+s^2 (4 (M_H^2-q_2) (q_1-q_2)\nonumber\\&&\;\;
+5 (q_1-3 q_2) s) t_k+s (5 M_H^4+6 q_1^2+5q_2 (q_2-5 s)+q_1 (-6 q_2+s)\nonumber\\&&\;\;
+M_H^2 (-6 q_1-4 q_2+4 s)) t_k^2+((M_H^2-q_2) (4 M_H^2-3q_1-q_2)+3 (3 M_H^2+2 q_1-5 q_2) s\nonumber\\&&\;\;
+s^2) t_k^3+(M_H^2-q_2+4 s) t_k^4) u_k-(-8(M_H^2-q_1) (q_1-q_2) s^2+s (4 (M_H^2-q_1) \nonumber\\&&\;\;
\times (M_H^2-q_2)+(-5 M_H^2+8 q_1-15 q_2) s) t_k+((M_H^2-q_1)(4 M_H^2-3 q_1-q_2)+(M_H^2\nonumber\\&&\;\;
-q_1-20 q_2) s+5 s^2) t_k^2+2 (M_H^2+2 q_1-4 q_2) t_k^3+t_k^4)u_k^2+(-4 (M_H^2-q_1)^2 s\nonumber\\&&\;\;
+(3 M_H^2-3 q_1+10 q_2) s t_k+(-5 M_H^2+q_1+10 q_2+s) t_k^2+t_k^3)u_k^3+4 q_2 t_k u_k^4
\Big]\;,\\
&&h(s,q_1,q_2,t_k,u_k)=\f{1}{t_k^2 q_2}
\Big[
2 (-M_H^4 t_k^2-q_2^2 (s^2+s t_k+t_k^2)+M_H^2 t_k (-M_H^2+t_k) u_k\nonumber\\&&\;\;
-(M_H^2-t_k)^2u_k^2-q_1^2 (s+u_k)^2+q_1 (s (-M_H^2 t_k+q_2 (2 s+t_k))+(2 (M_H^2+q_2) s\nonumber\\&&\;\;
+(M_H^2-q_2-2s) t_k) u_k+2 (M_H^2-t_k) u_k^2)+q_2 (-t_k (s^2+u_k (t_k+u_k))\nonumber\\&&\;\;
+M_H^2(s (t_k-2 u_k)+t_k (2 t_k+u_k))))
\Big]\;.
\eeeq

\bibliography{biblio}

\end{document}